\documentstyle[aps,prl,epsfig,floats,twocolumn]{revtex}
\def\be{\begin{equation}}
\def\ee{\end{equation}}

\def\H{{\mathcal{H}}}

\def\ba#1{\begin{array}{#1}}
\def\ea{\end{array}}
\def\G{\Gamma}
\def\bn{\begin{enumerate}}
\def\en{\end{enumerate}}
\def\sph{spin-$\frac{1}{2}$}
\def\spo{spin-$1$}
\def\spt{spin-$\frac{3}{2}$}

\title{Spin Reduction Transition in Spin-$\frac{3}{2}$ Random  Heisenberg Chains}
\author{{\sc Gil Refael,${}^1$ Stefan Kehrein,${}^2$ Daniel S. Fisher${}^1$} \\ 
{\small \em ${}^1$Dept.\ of Physics, Harvard University, Cambridge MA, 02138\\
${}^2$Theoretische Physik III, Elektronische Korrelationen und Magnetismus, Universit\"at Augsburg, Augsburg, Germany}}

\begin{document}
\wideabs{
\maketitle
\begin{abstract}
Random \spt\   antiferromagnetic Heisenberg chains are investigated using an  asymptotically exact renormalization group. Randomness is found to induce a quantum phase transition between two random-singlet phases. In the strong randomness phase the effective spins at low energies are $S_{eff}=\frac{3}{2}$, while in the weak randomness phase the effective spins are $S_{eff}=\frac{1}{2}$. Separating them is a quantum  critical point near which there is a non-trivial mixture of \sph,~\spo, and \spt\ effective spins at low temperatures. 
\end{abstract}


}


Some of the most dramatic effects of randomness in solids appear in the low temperature behavior of quantum systems.   A (deceptively) simple class of such systems are random quantum spin chains, in particular Heisenberg antiferromagnetic chains with Hamiltonian
\be
\H=\sum_i J_i {\bf \hat{S}_i}\cdot {\bf \hat{S}_{i+1}} \label{HAFM}
\ee
From a real-space renormalization-group (RG) analysis,\cite{MaDas} it has been shown that the \sph~random anti-ferromagnetic (AFM) chain is strongly dominated by randomness at low temperatures even when the disorder is weak.\cite{DSF94}  Its ground state is a  {\it random singlet} (RS)  phase in which pairs of spins --- mostly close together but occasionally arbitrarily far apart --- form singlets.  As the temperature is lowered, some of these singlets form at temperatures of order the typical exchange and become inactive. But their neighboring spins  will interact weakly across them via virtual triplet excitations. At lower temperatures, such further neighbors can form  singlets and  the process repeats. Concomitantly, the distribution of effective  coupling strengths broadens rapidly. Eventually, singlets form on all length scales and the ground state is controlled by an RG fixed point with extremely strong disorder: an {\it infinite randomness fixed point}.

This low temperature behavior is in striking contrast to that of the pure spin-$\frac{1}{2}$ AFM chain in which spin-spin correlations decay as $x^{-1}$ because of long-wavelength low-energy spin-wave (or spinon) modes. In the random singlet phase,   the {\it average correlations}  decay as a power of distance   --- as $x^{-2}$  --- but for a very different reason: A typical pair of widely spaced spins will have only exponentially (in the square root of their separation) small correlations. But a small fraction, those that form a singlet pair, will have correlations of order unity independent of their separation; these rare pairs completely dominate the average correlations as well as the other low temperature properties of the random system.

Infinite randomness fixed points are ubiquitous in random quantum systems. They probably control phase discrete-symmetry breaking transitions in {\it all} random quantum systems --- in any dimension ---  \cite{2DRQI} and, in addition to  the spin-$\frac{1}{2}$ AFM chain, also control the low-temperature properties of a range of  random quantum phases. \cite{OTHER-RQ-PHASES}  Because of their ubiquity,  further investigation of what types of random quantum phases and transitions can occur should shed light more generally on the combined roles of randomness and quantum fluctuations.  The simplest cases to analyze are one-dimensional beacuse asymptotically exact RG's can be used to extract much of the universal low-temperature behavior.  In this paper, we study random spin-$\frac{3}{2}$ AFM chains and find that they exhibit a novel phenomenon:  a quantum transition between two phases with both phases and the transition governed by infinite randomness fixed points. 

\begin{figure}
\psfig{figure=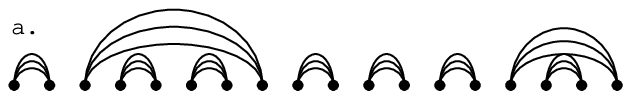,width=8.5cm}
\psfig{figure=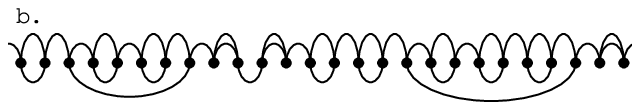,width=8.5cm}
\caption{Each connecting line represents a {\it spin-half singlet} link.  (a) Strong randomness \spt\      random-singlet phase. (b) Low randomness phase of a \spt\      chain: valence-bond solid + \sph\ random singlet.
 \label{phases}}
\end{figure}

We first review what is known about random spin-1 chains. 
Pure spin-$1$ AFM chains behave strikingly differently than spin $\frac{1}{2}$: their ground state is a non-degenerate disordered phase with excitations separated from it by a gap. \cite{haldane} This Haldane gap  provides robustness of spin-$1$ chains against {\it weak bounded} randomness.\cite{MGJ,HymanYang} But  for strong randomness, \spo\ chains will form a  random singlet phase. As is the case in many random quantum systems, there is {\it not} a transition directly from the gapped phase to the strong randomness phase. Instead,  when in some local regions the randomness overcomes the gap, there will be an intervening region in which there are localized gapless excitations but still exponential decay of correlations --- a Griffiths-McCoy phase. The system undergoes a quantum transition from this to the random singlet phase as the randomness is increased further.\cite{MGJ,HymanYang}

Pure \spt\   chains with Heisenberg interactions are gapless and behave very much like their \sph\   counterparts. \cite{pure3/2,LiebSchulzMattis}  We will show that {\it random} \spt\    chains undergo a phase transition as a function of the randomness between two zero-temperature phases:  the strong disorder phase is the \spt\   analog of the RS phase, with pairs of spins forming singlets [Fig. \ref{phases}(a)]. Surprisingly, the weak randomness phase is also an RS  phase, but of an effectively  \sph\   chain  superimposed on  a Haldane phase [Fig. \ref{phases}(b)]. At a critical disorder, there is a transition between these phases,  with special  behavior at the critical point, including a specific combination of \sph, \spo, and \spt\   character at low temperatures.

\begin{figure}
\resizebox{8.5cm}{5cm}{\psfig{figure=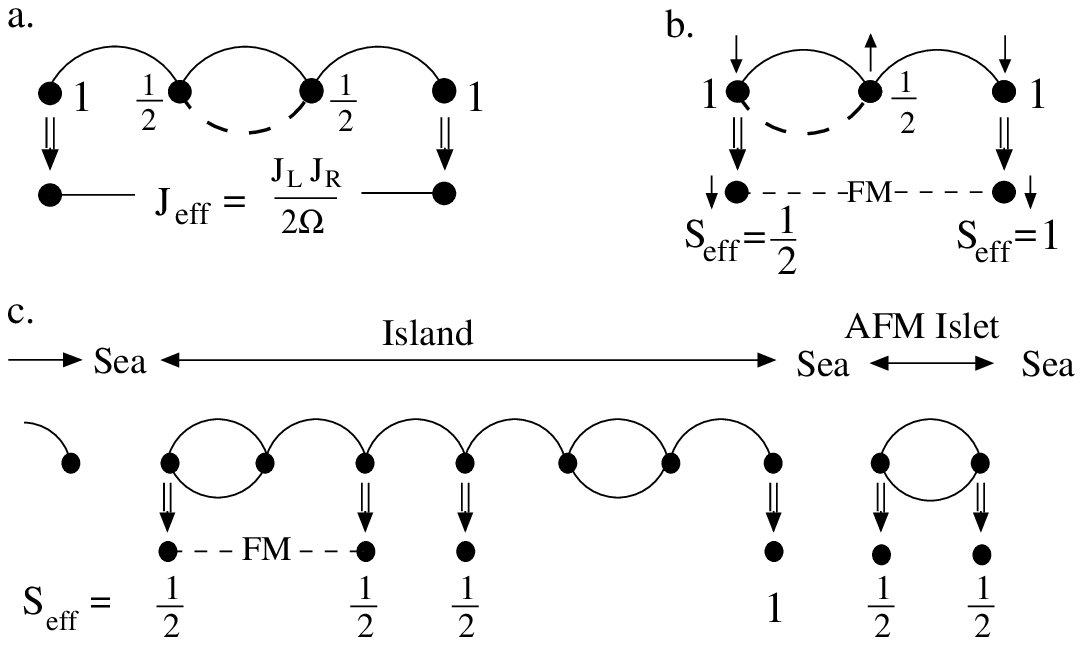,width=8.5cm}}
\vspace{1mm}
\caption{(a) RG rule for the final decimation of two spin $\frac{3}{2}$'s; these are connected by a dashed line which represents a \sph~singlet link being formed. (b) Creation of a FM bond by the formation of a link marked by a dashed line. Small arrows indicate preferred relative orientation of the active spins. (c) Low energy structure of \spt~chain showing a valence bond solid {\it island}, composed of effective \sph's 
antiferromagnetically coupled  in its interior, with a \spo~or a \sph~pair ferromagnetically coupled at its ends, separated from other islands by AFM {\it sea} bonds.  An AFM {\it islet}, made of two spin $\frac{3}{2}$'s joined by two links, is also shown.
Solid arcs represent already formed links; the effective spin is noted next to each site.  
\label{MDandFM}}
\end{figure}

To make progress, we first review the  RG \cite{MaDas} analysis of random \sph\  AFM  chains. This proceeds by gradually reducing the energy scale, $\Omega$.  First, the pair of  spins with the strongest coupling,  $J_{max}=\Omega_I$  --- the initial energy scale --- forms  a singlet [Fig. \ref{MDandFM}(a)], and is decimated. Virtual triplet excitations cause the two sites neighboring the singlet to weakly interact with the effective coupling:


\be
J_{eff}\approx \alpha\frac{J_\ell J_r }{J_{max}}   \label{MD}
\ee
where $J_\ell ,\,J_r $ are, respectively, the bonds to the left and right of the decimated pair and $\alpha=\frac{1}{2}$. By repeating this procedure, we gradually reduce the energy scale, $\Omega$, and the number of active spins in  the chain.  In the limit of low energy, the random singlet phase emerges and singlets form on all length scales. That this occurs for arbitrarily weak randomness, as it does, \cite{DotyFisher} cannot be convincingly shown by this RG as it is initially approximate when the distribution of $J$'s is not broad.  But its qualitative validity for weak randomness is suggested, since $J_{eff}$ is always less than $J_{\ell,\,r}$ due to the prefactor $\frac{1}{2}$ in Eq. (\ref{MD}). The multiplicative structure of Eq. (\ref{MD}) suggests that the distribution   of $J$'s broadens without bound. This means that the perturbative result (\ref{MD}) becomes {\it exact} at late stages of the RG, \cite{DSF94} and the universal low energy properties of the system can be found exactly.
  
The wide distribution of $J$'s allows one to associate the renormalized energy scale $\Omega$ with the temperature $T$. Bonds stronger than $T$ become frozen, and the remaining spins act as though they are free since almost all of their couplings are much weaker than $T$ at low temperatures. 

The RG flow is simply parametrized in terms of
\be
\ba{cc}
\Gamma=\ln\frac{\Omega_I}{\Omega},\, & \,\beta_i=\ln\frac{\Omega}{J_i} \ . \label{log}
\ea
\ee 
As the RG evolves, $\Omega$ is reduced, and $\Gamma$ increases. 
At low energies the coupling distributions  become scale invariant functions of $\beta/\G$; as $\G\rightarrow \infty$ at the fixed point,  the distributions  become infinitely broad. 
The density of active spins decays as
\be
\rho\sim\frac{1}{\G^{1/\psi}} \label{ndecay}
\ee
with $\psi=\frac{1}{2}$ a universal exponent characterizing the  random singlet phase.\cite{DSF94} As $\psi$ relates the {\it logarithm} of energy scales to length scales ($1/\rho$), it replaces the  exponent $z$ which parametrizes power-law energy-length scaling at conventional quantum critical points.  

The strong randomness phase of the \spt~chain can be understood similarly.  Combining strongly interacting neighbors into a singlet  yields Eq. (\ref{MD}) with $\alpha=\frac{5}{2}$.   
Strong randomness in the $J$'s will guarantee that despite the large prefactor ($\frac{5}{2}$) the new coupling will almost always obey $J_{eff}<J_{\ell,\,r}$, yielding flow towards  the random-singlet phase. In Fig. \ref{phases}(a), this is indicated by  varying length {\it triple links} representing singlets of spin $\frac{3}{2}$.

When the randomness is weak, the RG for spin $\frac{3}{2}$~fails to reduce the energy scale, suggesting that strong randomness behavior might not be obtained. To proceed, we  generalize the method of Monthus, Golinelli and Joliceur.\cite{MGJ} Instead of fully decimating strongly coupled pairs of spins, we only {\it partially} decimate  them,  eliminating their highest energy subspace.  Thus, when a spin pair, ${\bf S}_L,\ {\bf S}_R$,  is renormalized, its totally ferromagnetic (maximum spin) combination is eliminated. This corresponds to  breaking each spin into spin-$\frac{1}{2}$ parts ---  a \spt~consists of three \sph's  symmetrized --- with each contributing one \sph~to form a {\it spin-half singlet link}, leaving  a pair of spins with  $S'_{L,\ R}=S_{L,\ R}-\frac{1}{2}$ and modified couplings between them as well as between each one and its other neighbor. In the ground state, every site  must have three links joining it to others, e.g.,  as in Fig. \ref{phases}. When a link forms between two \sph's [Fig. \ref{MDandFM}(a)], both spins disappear and the $J_{eff}$ between the remaining neighboring spins is given by Eq. (\ref{MD}). As can be seen in Fig. \ref{MDandFM}(b), whenever only {\it one} of an antiferromagnetically coupled pair is spin $\frac{1}{2}$, it  will be decimated, and its partner will form a {\it ferromagnetic effective bond} across it.  Such ferromagnetic (FM) bonds can themselves be decimated forming, e.g.,  a  spin $\frac{3}{2}$    from a \spo\   and \sph\   pair; however, no spins greater than $\frac{3}{2}$ can form.  
We thus see, that as the energy scale is lowered, the {\it distribution of effective spins} changes.  In the strong randomness phase, at low energies virtually all the active (undecimated) spins   have $S_{eff}=\frac{3}{2}$. But this will not be the case when the randomness is weak.

For a \spo~chain with a narrow distribution of exchanges,  i.e.~weak randomness, all of the bonds between \spo 's would rapidly be partially decimated.  The resulting (approximate)  state which has one link connecting each site with each of its neighbors is the  {\it valence bond solid} picture of the Haldane phase.\cite{AKLT} The scale $\Omega_\infty$ at which the last spin is eliminated is the gap.   For stronger randomness, some double links will form and the gap will disappear.  But not until a critical randomness is reached does the continuous line of links break into finite segments; it is this that distinguishes the topological order of the Haldane phase from the random singlet phase.\cite{MGJ,HymanYang}

The phases of a \spt~chain can be understood in a related way. With weak randomness, decimation induces singlet links between most neighboring pairs, creating {\it islands} of valence-bond solid. Inside the islands, the active degrees of freedom are spin $\frac{1}{2}$'s left over from the decimations with spin $1$'s at the ends of islands [Fig. \ref{MDandFM}(c)]. The islands grow until the entire chain consists of one island with only spin $\frac{1}{2}$'s remaining. At lower energies, these spin $\frac{1}{2}$'s form \sph~random singlets: the ground state is thus a \sph~random singlet phase superimposed on a (\spo-like) valence bond solid; see  Fig. \ref{phases}(b).


Generally, the low energy structure of a \spt~chain will consist of valence bond islands separated  by AFM ``sea" bonds with no links yet formed across them.  Each island  consists of  a number --- possibly zero ---  of antiferromagnetically coupled active \sph's in the interior with each end being  either \spo\  or two ferromagnetically coupled \sph's as in Fig. \ref{MDandFM}(c). The exceptions to this are {\it islets}   consisting of a single AFM bond between two \sph~ends; these  arise  from  a pair of \spt~sites connected by two links [Fig. \ref{MDandFM}(c)]. There can also be original undecimated spin $\frac{3}{2}$'s.

It is convenient to describe all this in terms of  a purely {\it \sph~effective model} with a spin $1$~represented as a pair of \sph~sites with a FM interaction stronger than the energy scale, $\Omega$, and a spin $\frac{3}{2}$~by an island of three sites with two strong FM bonds. This has the advantage that coupling distributions and bond types remain {\it independent} if they are so initially; thus the number, $n$,  of internal \sph's in an island is distributed {\it exponentially} with density $\propto B^n$.  There are four coupling distributions: AFM sea bonds, FM edge bonds, AFM intra-island bonds, and (AFM) islet bonds. The other parameters are $B$ and $q$, the fraction of active spins that are in islets. The RG flows always broaden without bound the distributions of weak ($<\Omega$) bonds,  \cite{RKFlong} justifying the claim that the RG  is asymptotically exact.  
In the strong randomness limit, $B\to 0,\ q\to0$, so that all islands are three spin $\frac{1}{2}$'s strongly FM'ly coupled internally and weakly AFM'ly coupled between them equivalent to spin $\frac{3}{2}$'s. In contrast, for weak randomness at low energies, $B\to 1$ and $q\to 0$, so that an infinite island forms and the system becomes equivalent to a random \sph~chain; this then forms a \sph\  RS phase. Separating these two zero-temperature phases is a novel critical point with non-trivial $B$ and $q$.  Both phases and the critical point are controlled by infinite randomness fixed points.

To verify the above claims and quantitatively study the critical point, we implemented the full RG numerically. Initially, $\H$ (\ref{HAFM}) is all spin $\frac{3}{2}$~with the $J$'s uniformly distributed in $(J_{min},\,J_{max})$ and we define  $\delta\equiv {\rm var}(\ln J)$. We studied 100 realizations of length $5\times 10^{6}$,  measuring the evolution with energy scale of the active spin density $\rho$,
the effective spin distribution, and the coupling distributions.

For $\delta>\delta_c$ the chain flows to the $S_{eff}=\frac{3}{2}$ random-singlet phase, while for $\delta<\delta_c$ it flows to the $S_{eff}=\frac{1}{2}$ random singlet phase. The density $\rho$ in both random singlet phases obeys Eq. (\ref{ndecay}), as expected, with $\psi=\frac{1}{2}$. 

The critical point is at $\delta_c=0.22\pm0.01$. The corresponding fixed point is very different from the stable fixed points. The fractions of  active spins are ($\pm0.02$):
\be
\ba{ccc}
p_{\frac{1}{2}}=0.54 \,\,& p_{1}=0.33\,\, & 
p_{\frac{3}{2}}=0.13 \label{ps}
\ea
\ee
The appearance of spin-1 excitations may be surprising: in pure \spt chains, they do  {\it not} appear at the ends because of the gapless nature of the bulk.  At the critical point, the active spin density, $\rho$, decays with a larger power of $\G$ than in either phase: 
\be
\frac{1}{\psi}=\frac{1}{\psi_c}=3.85\pm 0.15 \label{psic}.  
\ee
This implies that the dynamics is {\it faster} at the critical point than in the adjacent phases. At infinite randomness fixed points, $\psi$ also controls the decay of {\it typical} correlations: \cite{DSFreview,2DRQI} 
\be
\ln(|\langle {\bf S}_i\cdot {\bf S}_j\rangle|) \approx  - C_{ij} |i-j|^{\psi}
\ee
with the random coefficient $C_{ij}$ having a universal distribution. The {\it average} correlations will, however, decay as $1/|i-j|^2$ at the critical point as in both phases.

\begin{figure}
\resizebox{8cm}{5.7cm}{\psfig{figure=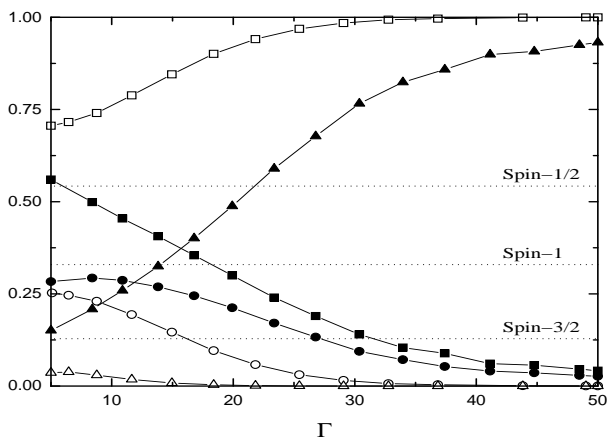,angle=0,width=8.5cm}}
\vspace{1mm}
\caption{Evolution of the effective spin fractions as a function of $\G$. Squares are spin 1/2, circles spin 1, and triangles spin 3/2. Filled symbols mark high randomness: $\delta=0.44$; empty symbols mark low randomness: $\delta=0.04$. The three horizontal lines mark the value of the fractions at the fixed point, $\delta=0.22$. \label{Seff}
}
\end{figure}

Deviations from the critical point, $\delta-\delta_c$,  are relevant perturbations and grow as $\G^{1/\nu\psi_c}$ as the energy scale is reduced, with $\nu$ the correlation length exponent. we find 
\be
\ba{cc}
\frac{1}{\nu\psi_c}=1.2\pm 0.1 & \hspace{3mm} \Rightarrow \hspace{3mm} \nu=3.2\pm 0.3 \ . \label{chi}
\ea
\ee


Many physical quantities are dominated by the almost decoupled active spins that remain at scale $\Omega=T$ corresponding to $\Gamma_T=\ln(\Omega_I/T)$.  The magnetization density at temperature $T$ and applied field $H \sim T$ is the the sum of that of the three kinds of spins with weights, $\{p_{S\,(\G_T,\,\delta)}\}$ (see, e.g., Fig. \ref{Seff}). The linear susceptibility obeys a universal scaling form: 
$\chi(\delta,T)\approx\frac{\rho_{\G_T}}{T}\sim \G_T^{-1/\psi_c}{\cal N}\left((\delta-\delta_{c}) \G_T ^{1/\psi_c \nu}\right)/T$.
For $x\rightarrow 0$, ${\cal N}(x)$ approaches a nonzero constant, yielding
$\chi(T)\approx 1/T\ln^{1/\psi_c} T$ for $|\delta-\delta_c|<|\ln T|^{-1/\psi_c\nu}$.
For large $x$,  ${\cal N}{(x)}\sim |x|^{(1-2\psi_c)\nu}$ leading to $\chi(\delta,T)\approx X(\delta)/T\ln^2 T$ 
in { \it both} random singlet phases.   
Near the critical point $X(\delta)$ vanishes as $X(\delta)\sim|\delta-\delta_c|^{(1-2\psi_c)\nu}$ 
for $|\delta-\delta_c|>|\ln T|^{-1/\psi_c\nu}$. Unfortunately, this dip in the susceptibility would be hard to observe because of the low temperatures needed. But because of the $\ln T$ in scaling functions, a wide regime of the low temperature phase diagram will be governed by the {\it critical} fixed point with the spin mixtures described approximately by the universal fractions in Eq. (\ref{ps}).

Spatiotemporal correlations can be investigated by neutron scattering.  The magnetic structure factor, ${\cal S}(q,\omega)$, will be dominated at low frequencies by excitations of spins that are paired together with energy scale $\omega$.  At fixed $\omega$,  ${\cal S}(q,\omega)$ will show a peak at $q\sim \rho_{\Gamma(\omega)}$,  the typical spacing between such spin pairs.  \cite{Motrunich-Damle} At the critical point, we also expect some strong {\it ferromagnetic correlations} between widely separated pairs on the same sublattice.  These may give rise to interesting dependence on $\delta$ of the  peak in ${\cal S}(q,\omega)$ near the zone boundary.  


The dynamics of nominally pure \spt~Heisenberg chains were recently studied  experimentally in $CsVCl_3$ and $CsVBr_3$, cf. Itoh et al.\cite{Itoh}
If mixtures of these, or other pairs of compounds can  be made with random AFM exchange, it should be possible to investigate some of the phenomena discussed here.  Additional complications that would have to be investigated include the effects of random anisotropy.  For \sph~random chains, there is considerable robustness of the random singlet-like phases unless Ising anisotropy dominates.\cite{DSF94} But for higher spin, this needs exploring.  Another intriguing possibility is three-leg spin ladder compounds, \cite{three-leg} if ones can be found with combinations of ferro and antiferromagnetic interactions. More generally, the model studied here shows how regimes with complicated mixtures of effective spins can arise at low temperatures from seemingly-simple hamiltonians.

The \spt~AFM chain appears to be the first example of a system in which two phases and the transition between them are all governed by infinite randomness fixed points.  How much of this behavior persists in other contexts, in particular with lower symmetry or in higher dimensions, is a subject for future investigations.  

\vspace{3mm}

This work has been supported in part by the National Science Foundation via grants DMR-9976621 and DMR-9809334. S.K. was also supported by a DFG Fellowship and through Grant No. SFB~484 of the DFG. G.R. would like to thank J. P. Sethna for useful discussions.

\end{document}